\documentclass[10pt]{article}

\usepackage{amsfonts} 
\usepackage{amsmath}    
\usepackage{amssymb}
\usepackage{a4wide}

\usepackage{latexsym} 
\newcommand{\ket}[1]{|#1\rangle} 
\newcommand{\inner}[2]{(#1,#2)} 
\newcommand{\ox}{\otimes} 

\newcommand{\C}[1]{\mathrm{#1}}
\newcommand{\many}{{\leq}_m^p}
\newcommand{\Tu}{{\leq}_T^p}
\newcommand{\qbf}{\textsc{qbf}}
\newcommand{\Tt}{{\leq}_{\mathit{tt}}^p}
\newcommand{\sat}{\textsc{sat}}
\newtheorem{theorem}{Theorem}
\newtheorem{corollary}{Corollary}
\newenvironment{proof}{\noindent \textbf{Proof:}}{\mbox{$\Box$}}

\newcommand{\pair}[1]{\langle #1 \rangle}
\newcommand{\lng}[1]{|#1|}
\newcommand{\sortkey}[1]{} 

\title{\textbf{Quantum Bounded Query Complexity}} 
\author{\emph{Harry Buhrman}\\
Quantum Computing and Advanced Systems Research, C.W.I. 
Amsterdam\thanks{Quantum Computing 
and Advanced Systems Research, C.W.I., P.O. Box 94079, NL--1098~GB 
Amsterdam, The Netherlands.} \\
\texttt{buhrman@cwi.nl}
\and
\emph{Wim van Dam}\\
Centre for Quantum Computation,
University of Oxford\thanks{Centre for Quantum Computation, 
Clarendon Laboratory, University of Oxford, Parks Road, Oxford, OX1~3PU,
United Kingdom.}\\
Quantum Computing and Advanced Systems Research, C.W.I. Amsterdam$^*$\\
\texttt{wimvdam@qubit.org}}

\begin{document} 
\maketitle 
\begin{abstract}
We combine the classical notions and techniques for
bounded query classes with those developed in quantum computing.  
We give strong evidence that quantum queries to an oracle
in the class $\C{NP}$ does indeed reduce the query complexity
of decision problems. Under traditional complexity assumptions, 
we obtain an exponential speedup between the quantum and the classical
 query complexity of function classes.

For decision problems and function classes 
we obtain the following results:
\begin{itemize}
\item{$\C{P}^{\C{NP}[2k]}_{||} \subseteq \C{EQP}_{||}^{\C{NP}[k]}$} 
\item{$\C{P}_{||}^{\C{NP}[2^{k+1} -2]} \subseteq \C{EQP}^{\C{NP}[k]}$}
\item{$\C{FP}^{\C{NP}[2^{k+1}-2]}_{||} \subseteq \C{FEQP}^{\C{NP}[2k]}$}
\item{$\C{FP}^\C{NP}_{||} \subseteq \C{FEQP}^{\C{NP}[O(\log n)]}$}
\end{itemize}
For sets $A$ that are many-one complete for
$\C{PSPACE}$ or $\C{EXP}$ we show that 
$\C{FP}^A \subseteq \C{FEQP}^{A[1]}$.  
Sets $A$ that are many-one complete for $\C{PP}$ have the property that
$\C{FP}_{||}^A \subseteq \C{FEQP}^{A[1]}$. In general    we prove that
for any set $A$ there is a set $X$ such that $\C{FP}^A \subseteq
\C{FEQP}^{X[1]}$, establishing that no set is superterse in the quantum
setting.
\end{abstract} 
\section{Introduction} 
The query complexity of a function is the minimum number of queries 
(to some oracle) that are needed to compute one value of this function.
With \emph{bounded} query complexity we look at the set of functions that
can be calculated if we put an upper bound on the number of queries
that we allow the computer to ask the oracle.
This notion has been 
extensively studied both in the resource bounded setting
~\cite{AgrawalArvind96,AmirBeigelGasarch90,AmirG88,Beigel87a,Beigel88,Beigel91,BeigelKummerStephan95,Kadin88,Ogihara95,Wagner90}   
and in the recursive setting\cite{BeigelGasarch89,BeigelGasarchGillOwings93}. 
This notion and its variants has lead to a series of techniques and tools 
that are used throughout complexity theory.  
 
In this paper we combine some of the bounded query notions with
quantum computation.
 The main goal of the paper is to further---as was 
done by Fortnow and Rogers~\cite{FortnowRogers98}---the
incorporation of quantum computation into complexity theory. We feel 
that the synthesis of quantum computation and classical complexity 
theory serves two purposes. First, it is important to know the limits of 
feasible quantum computation and this can be done by incorporating it 
into the framework of classical computation. Second, the insights of 
quantum computation can be useful for classical complexity theory
in turn.

We start out with the class of sets (or decision problems) that are 
computable in polynomial time with bounded queries to a set in $\C{NP}$.
We consider the setting where the queries are adaptive 
(i.e., a query may depend on the answers to previous ones), 
as well as where they are non-adaptive.  Classically, it is 
known that any decision problem that can be solved in polynomial time 
with $k$ adaptive queries to a set in  $\C{NP}$
(the class $\C{P}^{\C{NP}[k]}$) can also be solved
with $2^k-1$ non-adaptive queries 
(the class $\C{P}_{||}^{\C{NP}[{2^k}-1]}$,
where ``$||$'' indicates the parallel or non-adaptive queries),
and vice-versa~\cite{Beigel91}.
In other words:
$\C{P}^{\C{NP}[k]} = \C{P}^{\C{NP}[2^k-1]}_{||}$.
Moreover, there is strong evidence that this trade-off is optimal
in the sense that every non-adaptive class $\C{P}_{||}^{\C{NP}[k]}$ 
is different for different values of $k$.
For example if $\C{P}^{\C{NP}[2]}_{||} \subseteq \C{P}^{\C{NP}[1]}$, 
then the  polynomial hierarchy collapses~\cite{Kadin88}
(see also~\cite{BuhrmanFortnow98,HemaspaandraHemaspaandraHempel97}).  
 
The natural quantum analogue of $\C{P}$ is the class $\C{EQP}$, which
stands for \emph{exact quantum polynomial time.} This is the class of
sets or decision problems that is computable in polynomial time with a
quantum computer that makes no errors (i.e., is exact).  In this paper
we will see that if we allow the query machine to make use of quantum
mechanical effects such as superposition and interference the
situation changes. In the non-adaptive case we will show that $2k$
classical queries can be simulated with only $k$ non-adaptive ones on
a quantum computer and in the adaptive case we show how to simulate
$2^{k+1}-2$ classical queries with only $k$ quantum queries.  Hence
\begin{eqnarray*}
\C{P}_{||}^{\C{NP}[2k]} \subseteq \C{EQP}^{\C{NP}[k]}_{||}
& \mathrm{ and } &
\C{P}_{||}^{\C{NP}[2^{k+1}-2]} \subseteq \C{EQP}^{\C{NP}[k]}.
\end{eqnarray*} 
In particular it follows from this result that 
$\C{P}_{||}^{\C{NP}[2]} \subseteq \C{EQP}^{\C{NP}[1]}$
(see also~\cite{vanDam98}). 

In order to prove these  results  we combine the 
classical mind-change technique~\cite{Beigel91} with the one query version
(see \cite{CEMM98})
of the first quantum algorithm developed by  
David Deutsch~\cite{Deutsch85}. 
 
Next, we turn our attention to \emph{functions} that are 
computable with bounded queries to a set in $\C{NP}$. 
Compared to the decision
problems  there is probably no nice trade-off between
adaptive and non-adaptive queries for functions. 
This is because the following is 
known~\cite{BeigelKummerStephan95}: for any $k$ the inclusion 
$\C{FP}_{||}^{\C{NP}[k]} \subseteq 
\C{FP}^{\C{NP}[k-1]}$ implies that $\C{P} = \C{NP}$. 
Moreover, if $\C{FP}^{\C{NP}}_{||} \subseteq \C{FP}^{\C{NP}[O(\log n)]}$ 
then the polynomial time hierarchy collapses~\cite{Beigel88,Selman94,Toda91}. 
 
When the adaptive query machine is a quantum computer, things are different 
and we seem to get a trade-off between adaptiveness and query complexity. 
We show the following:
\begin{eqnarray*}  
\C{FP}_{||}^{\C{NP}[2^{k+1}-2]} \subseteq \C{FEQP}^{\C{NP}[2k]}  
& \mathrm{ and } &
\C{FP}_{||}^{\C{NP}} \subseteq \C{FEQP}^{\C{NP}[O(\log n)]}.
\end{eqnarray*} 
Here $\C{FEQP}^{\C{NP}[k]}$ is the class of 
functions that is computable by an exact quantum Turing machine that 
runs in polynomial time and is allowed to make $k$ queries to a set in 
$\C{NP}$.  The proofs of these results use our previous results 
on decision problems and a quantum algorithm developed by 
Deutsch-Jozsa~\cite{DeutschJozsa92} and 
Bernstein-Vazirani~\cite{BernsteinVazirani97}. 
 
Using the same ideas we are able to show that for any set $A$ there 
exists a set $X$ such that $\C{FP}^A \subseteq \C{FEQP}^{X[1]}$, 
establishing that no set is `superterse'. Also because the complexity of 
$X$ is not much harder than that of $A$ 
(the problem $X$ is Turing reducible to $A$), 
we get quite general theorems for complete sets of complexity classes.  
 
For a complexity class $\mathcal{C}$ that is closed under Turing reductions,
and a problem $A\in \mathcal{C}$ that is many-one complete for the 
class $\mathcal{C}$, the inclusion
$\C{FP}^{\mathcal{C}} \subseteq \C{FEQP}^{A[1]}$ is proven.
This holds in particular for the set $\qbf$ of the \emph{true 
quantified Boolean formulae} which is a $\C{PSPACE}$ complete problem, 
and the complete sets for the class $\C{EXP}$.
If $\mathcal{C}$ is a class that is closed under
truth-table reductions, then it holds that
$\C{FP}_{||}^{\mathcal{C}} \subseteq \C{FEQP}^{A[1]}$.
The Theta levels of the polynomial hierarchy and $\C{PP}$ are
examples of such classes.
 
The ingredients for all our results are standard quantum algorithms 
combined with well known techniques from complexity theory. Nevertheless we 
feel that this combination gives a new point of view on the nature of 
bounded query classes and the structure of complete sets in general. 
 
\section{Preliminaries}\label{sec:prel} 
\subsection{Classical computing} 
We assume the reader to be familiar with basic notions of complexity theory 
such as the various complexity classes and types of reducibility
as can be found in many textbooks in the area
\cite{BalcazarDiazGabarro88,BalcazarDiazGabarro90,GareyJohnson79,Johnson94}. 
The essentials for this article are mentioned below.
 
For a set (decision problem) $A$ we will identify $A$ with its 
characteristic function. 
Hence for a string $x$ we have $A(x)\in\{0,1\}$, and $A(x)=1$ if and only 
if $x\in A$. A class $\mathcal{C}$ consists of a set of decision 
problems. A problem $A$ is many-one, or $\many$-complete 
for a class $\mathcal{C}$ if for any problem $B \in \mathcal{C}$, there
exists a polynomial function or ``Karp-reduction'' $\tau$ such 
that $x\in B$  if and only if $\tau(x) \in A$. The typical
example of such a complete problem is $\sat$ (the set of satisfiable Boolean
formulae) which is $\many$-complete for the class $\C{NP}$.
The class $\C{FP}$ indicates the set of \emph{functions} that can be calculated
on a polynomial time, deterministic Turing machine.

An oracle Turing machine is \emph{non-adaptive}, if it can produce a list of 
all of the oracle queries it is going to make before it makes the 
first query.  
For any set $A$, the elements of the class $\C{P}^{A[k]}$ ($\C{FP}^{A[k]}$) 
are the languages (functions) that are computable by polynomial time 
Turing machines that accesses the oracle $A$ at most $k$ times on each input. 
The class $\C{P}_{||}^{A[k]}$ and $\C{FP}_{||}^{A[k]}$ allow only 
non-adaptive access to $A$. 
The notation $\C{P}^{\C{NP}[q(n)]}$ is used to indicate algorithms
that might require $q(n)$ oracle calls, where $q$ is a function of the
input size $n$.

The class $\C{NP}$ can be generalised by defining the \emph{polynomial time 
hierarchy}. We start with the definition $\Sigma_0^p= \C{P}$ 
and then for the higher levels continue in an inductive fashion with  
$\Sigma_{i+1}^p=\C{NP}^{\Sigma_i^p}$ for $i=1,2,\ldots$ Many complexity 
theorists conjecture that this polynomial time hierarchy is infinite, 
i.e., $\Sigma_{i+1}^p\neq\Sigma_i^p$ for all $i$. 

A class $\mathcal{C}$ of languages is closed under Turing (truth-table)
reduction if any decision problem that can be solved with
a polynomial time Turing machine and (non-adaptive) queries to a set in $\mathcal{C}$,
is itself also an element of $\mathcal{C}$. Examples of such
classes are $\C{PSPACE}$, $\C{EXP}$, and the Delta levels
$\Delta^p_{i+1} = \C{P}^{\Sigma_i^p}$ of the polynomial time hierarchy. 
 The classes $\C{PP}$ and 
$\Theta^p_{i+1} = \C{P}_{||}^{\Sigma_i^p}$ (Theta levels of the polynomial
hierarchy) are for example closed under this truth-table-reduction.

\subsection{Quantum computing} \label{sec:quantcomp}
In this section we define quantum oracle Turing Machines.  
For an introduction to quantum computing see for example the survey by 
Berthiaume~\cite{Berthiaume97}. 
 
A \emph{qubit} is a superposition $\alpha_0\ket{0}+\alpha_1\ket{1}$ of 
both values of a classical bit. The complex values $\alpha_0$ and 
$\alpha_1$ are the \emph{amplitudes} of the quantum state,
and they obey the normalisation restriction: $|\alpha_0|^2+|\alpha_1|^2=1$.

The tensor or Kronecker-product is used to describe a system of several 
qubits.
The combination of two qubits is thus calculated by
\begin{eqnarray*}
\ket{x}\ox\ket{y} & = & 
\left({\alpha_0\ket{0} + \alpha_1\ket{1}}\right)
\ox
\left({\beta_0\ket{0}+\beta_1\ket{1}}\right) \\
& = & 
\alpha_0\beta_0\ket{00} +
\alpha_0\beta_1\ket{01} +
\alpha_1\beta_0\ket{10} +
\alpha_1\beta_1\ket{11}.
\end{eqnarray*}
Consequently, a register of $n$ qubits is a superposition $\ket{\psi}$ 
of all $2^n$ strings of $n$ classical bits, written 
\begin{eqnarray*}
\ket{\psi} & = & \sum_{i\in\{0,1\}^n}\alpha_i\ket{i}. 
\end{eqnarray*}
If we measure the quantum register $\ket{\psi}$ in the standard 
(i.e. classical) basis,
 we will observe one and only one of the basis states $\ket{i}$
with probability $|\alpha_i|^2$. 
Hence we must have $\sum_{i\in\{0,1\}^n}|\alpha_i|^2=1$ 
(the normalisation restriction). 
After measuring $\ket{\psi}$ and observing $\ket{i}$, we say that 
the superposition $\ket{\psi}$ ``has collapsed'' to the new 
state $\ket{i}$. 
 
If we do not observe a state, quantum mechanics tells us that it will evolve 
\emph{unitarily}, as this is the only evolution that respects the 
normalisation restriction.
Unitarity means that the vector of amplitudes is transformed according 
to a linear operator that preserves the unit norm. This can be viewed
as a rotation in the complex, finite Hilbert space of dimension $2^n$.
 A unitary operator $U$ always has an inverse $U^{-1}$ 
which equals its conjugate transpose $U^\dagger$. 
 
An example of a one-qubit operation is the Hadamard transform $H$ 
which is defined by
\begin{eqnarray*} 
H\ket{0} & = & \frac{1}{\sqrt{2}}({\ket{0}+\ket{1}}) \\ 
H\ket{1} & = & \frac{1}{\sqrt{2}}({\ket{0}-\ket{1}}). 
\end{eqnarray*}
The effect of the $n$-fold tensor product of $H^{\ox n}$ on 
$n$ qubits is now calculated by the linearity of quantum mechanics:
\begin{eqnarray*}
H^{\ox n}\sum_{i\in\{0,1\}^n}\alpha_i\ket{i} 
& = & 
\sum_{i\in\{0,1\}^n}\alpha_i \cdot H^{\ox n}\ket{i}.
\end{eqnarray*}
This shows that the evolution of the superposition $\ket{\psi}$ is 
fully determined by the evolution of its basis states $\ket{i}$.
We will use this transformation $H^{\ox n}$ in 
Section~\ref{sec:function}.

A quantum Turing machine's transition function can be described by a 
 unitary matrix $M$ with complex entries. The computation of $t$ 
time-steps will then correspond to the application of the matrix product 
$M^t$ to the
 initial configuration $\ket{j}$. 
At the end of the computation we measure the state that has evolved in 
this manner and accept $j$ if some designated bit is $1$ 
and otherwise reject the input $j$. 
 
The class $\C{EQP}$ is defined as those sets that can be computed by 
a quantum Turing machine that runs in polynomial time and accepts 
every string $j$ with probability $1$ or $0$.  Likewise, we define the 
class of functions $\C{FEQP}$ as the class of functions that can be 
computed by some quantum Turing machine that runs in polynomial 
time. The output of the Turing machine may now be several bits. 
 
We model oracle computation as follows (see 
also~\cite{BennetBernsteinBrassardVazirani97}). An oracle Turing 
machine has some special query tape, and at some point in the computation 
the Turing machine may go into a special pre-query state  to make 
a query to the oracle set $A$. Suppose the query tape contains the 
state $\ket{i}\ket{b}$ ($i$ represents the query and $b$ is a bit). 
The result of this operation is that after the call 
the machine will go into a special state called the post-query state 
and that the query tape has changed into $\ket{i}\ket{A(i)\oplus b}$, where 
$\oplus$ is the \textsc{exclusive or}. 
We will denote this unitary operation by $U_A$. 
Note that $U_A$ only changes the contents of the special 
query tape $b$, and leaves all the other registers unchanged.  

As with classical oracle computation, we make the distinction between 
adaptive and non-adaptive quantum oracle machines. We call a quantum 
oracle machine non-adaptive if on every computation path a list of all 
the oracle queries (on this path) is generated before the first query 
is made.  
 
The class $\C{EQP}^{A[k]}$ are the sets recognised by an exact quantum
Turing machine that runs in polynomial time and makes at most $k$
queries to the oracle for $A$. Again, we define classes like 
$\C{EQP}_{||}^{A[q(n)]}$,
$\C{FEQP}^{A[q(n)]}$, and $\C{FEQP}_{||}^{A[q(n)]}$, for non-adaptive
decision, adaptive function, and non-adaptive function classes
respectively (with $q(n)$ a function that gives an upper bound on the
number of queries and $n$ the size of the input string).
 
\section{Decision Problems} 
In this section we will investigate the extra power that a polynomial 
time, exact quantum computer yields compared to classical deterministic 
computation when querying a set in the class $\C{NP}$.  
In the case of deterministic computation the following equality between 
adaptive and non-adaptive queries to $\C{NP}$ is well known. 
 \newpage
\begin{theorem}\cite{Beigel91,BussHay91,Wagner90}\label{thm:classical-relation-sets} 
  \begin{enumerate} 
  \item For any $k\geq 0$ we have $\C{P}_{||}^{\C{NP}[2^{k}-1]} = 
    \C{P}^{\C{NP}[k]}$.\label{item:mindchange}  
  \item For any polynomial $q(n)>1$ the equality $\C{P}_{||}^{\C{NP}[q(n)]} = 
    \C{P}^{\C{NP}[O(\log(q(n)))]}$ holds.  
  \end{enumerate} 
\end{theorem} 
\begin{proof} 
Both items are proved in a similar way which has two parts. 
The first part shows that computing a function in 
$\C{P}_{||}^{\C{NP}[2^{k}-1]}$  can be reduced to computing the \emph{parity} 
of $2^{k}-1$ other queries to $\C{NP}$. The second part then proceeds 
by showing that using binary search one can compute the parity of $2^k-1$ 
$\C{NP}$-queries with $k$ \emph{adaptive} queries to $\sat$. 
On the other hand, it is trivial to see that any computation with 
$k$ adaptive queries 
can be simulated exhaustively with $2^k-1$ non-adaptive oracle calls.
\end{proof} 
 
There is also strong evidence that the  above trade-off is  
tight (see \cite{BeigelChangOgihara93,Kadin88}). 
It follows for example that if 
$\C{P}_{||}^{\C{NP}[2]} = \C{P}^{\C{NP}[1]}$ then the polynomial hierarchy 
collapses~\cite{Kadin88}. (See~\cite{BuhrmanFortnow98} for the latest 
developments with respect to this question.)  
 
Perhaps surprisingly the situation changes when the query machine is 
quantum mechanical. David Deutsch~\cite{Deutsch85} developed a quantum 
algorithm to compute the parity of any two Boolean variables in 
\emph{one} query with higher probability than a classical (randomised) 
algorithm can. Cleve \emph{et al.}~\cite{CEMM98} showed how to make this 
procedure exact. 
 
\begin{theorem}\cite{CEMM98,Deutsch85}\label{thm:parity} 
Let $f: \{0,1\} \mapsto \{0,1\}$. There exists an exact quantum 
algorithm that computes the parity bit $f(0) \oplus f(1)$ with one 
query to the function $f$. 
This algorithm works in constant time. 
\end{theorem} 
\begin{proof} 
For simplicity we only describe what is happening to the states that 
get effected by the oracle query. Construct the following initial state: 
\begin{eqnarray} 
\ket{\mathrm{Initial}} & = &  \label{eq:initial} 
\frac{1}{2}(\ket{0}+\ket{1})\ox (\ket{0}-\ket{1}). 
\end{eqnarray} 
Next, make the only query to $f$ depending on the value of the first bit.
 Note that $f$ will thus be queried in 
superposition for both $f(0)$ and $f(1)$. Applying $f$ establishes the
following evolution on the two qubits:
\begin{eqnarray*} 
\ket{i}\ox\ket{b} & \longrightarrow & \ket{i}\ox\ket{b\oplus f(i)}. 
\end{eqnarray*} 
This results in the following outcome when applied to the initial state: 
\begin{eqnarray*} 
\frac{(-1)^{f(0)}}{2}(\ket{0}+\ket{1})\ox(\ket{0}-\ket{1}) 
& \textrm{if} & f(0) = f(1)\\ 
\frac{(-1)^{f(0)}}{2}(\ket{0}-\ket{1})\ox(\ket{0}-\ket{1}) 
& \textrm{if} & f(0) \neq f(1).  
\end{eqnarray*} 
Which means that if we apply a Hadamard transformation to the first register,
we obtain 
\begin{eqnarray*} 
\ket{\mathrm{Final}} & = & 
{(-1)}^{f(0)}\ket{f(0)\oplus f(1)}\ox{(\ket{0}-\ket{1})}.  \label{eq:phase}
\end{eqnarray*}
Hence observing the first bit yields the correct  answer 
$f(0)\oplus f(1)$.
\end{proof} 

Using this procedure we will now show that a quantum Turing machine 
can compute decision problems with half the number of non-adaptive 
queries.  
\begin{theorem} \label{thm:2kversusk}
For any $k\geq 0$ we have the inclusion 
$\C{P}_{||}^{\C{NP}[2k]} \subseteq \C{EQP}_{||}^{\C{NP}[k]}$. 
\end{theorem} 
\begin{proof} 
Without loss of generality we will assume that the queries are made 
to $\sat$, and that the predicate that is computable with $2k$ queries 
to $\sat$ is $f(x)$. Let $\psi_1$, $\psi_2$,\ldots, 
$\psi_{2k}$ be the 
queries that the computation of $f(x)$ makes. 
We will use the proof technique of 
Theorem~\ref{thm:classical-relation-sets} (also called 
mind-change technique) which enables us to compute $f(x)$ by calculating
the single bit 
$\sat(\phi_1) \oplus \cdots \oplus \sat(\phi_{2k})$. Here the
new formulae $\phi_1, \ldots, \phi_{2k}$ can be computed in polynomial 
time from $\psi_1, \dots , \psi_{2k}$, $f$, and $x$, but without 
having to consult $\sat$.
 
Next, we use Theorem~\ref{thm:parity} to compute the parity 
$\sat(\phi_i) \oplus \sat(\phi_{i+1})$ for odd $i$ ($1\leq i<2k$) 
with $k$ non-adaptive queries to $\sat$. 
Finally we compute the parity of these answers, thus obtaining the
necessary information for calculating $f(x)$. 
\end{proof} 
\begin{corollary}\label{cor:2in1}
$\C{P}_{||}^{\C{NP}[2]} \subseteq \C{EQP}^{\C{NP}[1]}$ (see \cite{vanDam98}). 
\end{corollary}  
We do not know whether this is tight. It would be interesting to either 
improve this result to 
$\C{P}^{\C{NP}[2]} \subseteq \C{EQP}^{\C{NP}[1]}$ or to 
show as a consequence of this that the polynomial time hierarchy collapses. 
 
Theorem~\ref{thm:classical-relation-sets} relates adaptive query 
classes to non-adaptive ones, thereby establishing an exponential gain 
in the number of queries ($2^k-1$ versus $k$ queries). We will now show 
how to use the Deutsch trick to do even slightly better than that in the 
quantum case. 
 
\begin{theorem}\label{thm:quantum-bin-search} 
$\C{P}_{||}^{\C{NP}[2^{k+1}-2]} \subseteq \C{EQP}^{\C{NP}[k]}$ for
any $k\geq 0$. 
\end{theorem} 
\begin{proof} 
The proof is by induction on $k$. For $k=1$ we have back the situation 
of Corollary~\ref{cor:2in1}. Let the predicate $f(x)$ be computable with 
$2^{k+1}-2$ non-adaptive queries  to $\sat$.  
As in the proof of Theorem~\ref{thm:2kversusk} we reduce the 
$2^{k+1}-2$ queries $\psi_i$ that $f(x)$ makes, to the calculation of the 
parity-bit $\sat(\phi_1)\oplus\cdots\oplus\sat(\phi_{2^{k+1}-2})$. 
Next, we construct $2^{k+1}-2$ new formulae 
$\chi_1,\ldots,\chi_{2^{k+1}-2}$ according to: 
\begin{eqnarray*} 
\chi_i   \mbox{ is satisfiable }  & \Longleftrightarrow  & 
|{\{\phi_1,\ldots,\phi_{2^{k+1}-2}\} \cap \sat}| \geq i.
\end{eqnarray*} 
The construction of each such $\chi_i$ can be done in polynomial time. 
To see this, consider the non-deterministic polynomial time Turing 
machine $M$ that on input $\pair{i,\phi_1,\ldots,\phi_{2^{k+1}-2}}$, 
accepts if and only if it can find for $i$ of the formulae a satisfying 
assignment. Cook and Levin~\cite{Cook71,Levin73} ---proving that $\sat$ 
is $\many$ -complete for $\C{NP}$--- showed that any polynomial time 
non-deterministic Turing machine computation $M(x)$ in polynomial time 
can be transformed into a formula that is satisfiable if and only if 
$M(x)$ has an accepting computation. Let  $\chi_i$ be the result of 
this Cook-Levin reduction. 
 
Note the following two properties of those formulae $\chi_i$:
\begin{enumerate} 
\item{The parity $\sat(\phi_1)\oplus\cdots\oplus\sat(\phi_{2^{k+1}-2})$ 
is the same as the parity 
$\sat(\chi_1)\oplus\cdots\oplus\sat(\chi_{2^{k+1}-2})$.} 
\item{For every $i$ we have $\sat(\chi_i) \geq \sat(\chi_{i+1})$.
\label{monotone}}
\end{enumerate}
Now we are ready to make the first query. We compute the parity of 
$\chi_{2^{k-1}}$ and $\chi_{2^{k-1} + 2^{k} -1}$. This can be 
done in one query using Theorem~\ref{thm:parity}. By doing this we have 
at the cost of one query  reduced the question of computing the parity 
of $2^{k+1}-2$ formulae to computing the parity of $2^k-2$. These we 
can solve using $k-1$ queries using the induction hypothesis. 
To see this observe the following. For convenience set $a = 2^{k-1}$ 
and $b = 2^{k-1} + 2^{k} -1$.  
 
Suppose the parity of $\chi_a$ and $\chi_b$ is odd. Hence 
$\chi_1$, $\chi_2$,\ldots, $\chi_a$ are all satisfiable and 
$\chi_b,\ldots,\chi_{2^{k+1}-2}$  all un-satisfiable (using 
property~\ref{monotone} above). Also note that $a$ is even, so the 
parity of $\chi_1,\ldots,\chi_{2^{k+1}-2}$ is the same as the parity 
of $\chi_{a+1},\ldots,\chi_{b-1}$ (these are $2^k-2$ many formulae). 
 
On the other hand assume that the parity of $\chi_a$ and $\chi_b$ is 
even. This means (again using property~\ref{monotone} above) that 
$\chi_a,\ldots,\chi_b$ are all either satisfiable or un-satisfiable 
and hence have even parity. So again the question reduces to the 
parity of the remaining formulae: $\chi_1,\ldots,\chi_{a-1}$ and 
$\chi_{b+1},\ldots,\chi_{2^k+1-2}$. Which happen to be $2^k-2$ many 
formulae. 
\end{proof}

We do not know if it is possible to do better than this.  In essence
the above technique seems to boil down to searching in an ordered
list. Buhrman and De Wolf~\cite{BuhrmandeWolf98} show that this can
not be done faster than $\sqrt{\log n}/\log(\log n)$, which was
improved by Fahri \emph{et al.}~\cite{FarhiGGS98} to
$\log(n)/\ 2\log(\log(n))$ and later by Ambainis \cite{Ambainis99} to a
lower bound of $1/12 \log n - O(1)$ queries.  Recent results by Farhi
\emph{et al.} \cite{FarhiGGS99} seem to suggest a reduction in the
query complexity by a factor of two.  But it is not clear if their
exact quantum algorithm for `insertion into an ordered list'
translates correctly into our setting.
\section{Function Classes} \label{sec:function} 
Now we turn our attention to function classes where the algorithm 
can output bit \emph{strings} rather than single bits.  We will see
that in this scenario the difference between classical and quantum
computation becomes more eminent.

\subsection{Functions computable with queries to an oracle in NP} 
We start out by looking at functions that are computable with 
queries to a complete set for the class $\C{NP}$. 
Classically the 
situation is not as well understood as the class of decision 
problems. There is strong evidence that the analogue of 
Theorem~\ref{thm:classical-relation-sets} is not true. 
\begin{theorem}\label{thm:classical-functions} 
The following holds for the classical, exact computation of functions:
\begin{enumerate} 
\item If for some $k\geq0$ we have  
$\C{FP}_{||}^{\C{NP}[k+1]} \subseteq \C{FP}^{\C{NP}[k]}$, 
then $\C{P} = \C{NP}$~\cite{BeigelKummerStephan95}. 
\item If for all polynomials $q(n)$ (with $n$ the size of the input string): 
$\C{FP}_{||}^{\C{NP}[q(n)]} 
\subseteq \C{FP}^{\C{NP}[O(\log n)]}$, then $\C{NP} = \C{R}$ 
(and the polynomial hierarchy collapses)~\cite{Beigel88,Selman94,Toda91}. 
\end{enumerate} 
\end{theorem} 
When we allow the adaptive query machine to be quantum mechanical the
picture becomes again quite different.  We will show for example that
the inclusion $\C{FP}_{||}^{\C{NP}[q(n)]} \subseteq
\C{FEQP}^{\C{NP}[2\log(q(n))]}$ holds.  In order to do so we will
first need a generalisation of the parity trick of the 
Theorem~\ref{thm:parity}.
 
Deutsch and Jozsa~\cite{DeutschJozsa92} generalised the setting to 
that of
Boolean functions $f: \{0,1\}^n \mapsto \{0,1\}$ with the promise that 
 $f$ has  either an equal number of $0$ and $1$ outputs---they called 
such functions \emph{balanced}---or $f$ is \emph{constant} on all inputs
($0$ or $1$).  Classically one needs $2^n/2 + 1$ applications of $f$ to 
determine whether it is constant or balanced. Deutsch and Jozsa 
however showed that this can be done with $2$ applications in the 
quantum setting. Cleve \emph{et al.}~\cite{CEMM98} demonstrated how 
to do this with a single quantum query.  Based upon the work of Deutsch and 
Jozsa, Bernstein and Vazirani~\cite{BernsteinVazirani97} considered 
a subclass of the 
constant and balanced functions. Suppose that $f: \{0,1\}^n \mapsto 
\{0,1\}$ is such that: 
\begin{eqnarray}\label{eq:special-f} 
 f(x) & = & (a_1 \wedge x_1)\oplus\cdots\oplus (a_n\wedge x_n) = 
\inner{a}{x}. 
\end{eqnarray} 
Where $(a_1a_2\cdots a_n) = a \in \{0,1\}^n$, 
and $\inner{a}{x}$ is the inner product of the vectors $a$ and $x$ modulo $2$. 
The goal is, given $f$, to determine $a$. Bernstein and Vazirani showed 
that this can be done with $2$ applications of $f$ on a quantum 
computer. Classically one needs at least $n$ applications. 
Cleve \emph{et al.}~\cite{CEMM98} improved this again to only $1$ application 
of the function $f$. 
 
The core of the algorithm is the following observation. 
The $n$ fold Hadamard transform $H^{\ox n}$ (see Section~\ref{sec:quantcomp}) 
does the following when applied to a basis state of $n$ bits: 
\begin{eqnarray} \label{eq:hadamard-on-basis}
H^{\ox n} \ket{a_1 a_2 \cdots a_n} & = & 
\frac{1}{\sqrt{2^n}} 
\sum_{x\in\{0,1\}^n}{{(-1)}^{\inner{x}{a}}\ket{x}}. 
\end{eqnarray} 
Since the Hadamard transform is its own inverse we have also the other 
direction: 
\begin{eqnarray} \label{eq:hadamard-on-superposition}
 H^{\ox n} \frac{1}{\sqrt{2^n}} 
\sum_{x\in\{0,1\}^n}{{(-1)}^{\inner{x}{a}}\ket{x}} & = & 
\ket{a_1 a_2 \cdots a_n}.
\end{eqnarray} 
So, if we are able to obtain the state of
Equation~\ref{eq:hadamard-on-basis}, then we can \emph{extract} 
the $n$-bit string $a$ out of it by simply applying $H^{\ox n}$ to it.  
This state however can 
be obtained with one application to $f$ as follows: 
\begin{eqnarray}\label{eq:application-function} 
U_f \frac{1}{\sqrt{2^{n+1}}} \sum_{x \in \{0,1\}^n}{\ket{x}
(\ket{0} - \ket{1})}
&= &  \frac{1}{\sqrt{2^{n+1}}}\sum_{x\in\{0,1\}^n}{{(-1)^{f(x)}}\ket{x} 
  (\ket{0} - \ket{1})}.
\end{eqnarray} 
Now observe that the last qubit is always in state $(\ket{0} - 
\ket{1})$. Using the definition of $f$ we can rewrite this state to:  
\begin{eqnarray*} 
& &   \frac{1}{\sqrt{2^{n+1}}}\sum_{x\in\{0,1\}^n}{{(-1)}^{\inner{x}{a}}
\ket{x}}\ox 
(\ket{0} - \ket{1}). 
\end{eqnarray*}
Let us turn back now to our setting of bounded query classes. Using the 
above quantum tricks we can show the following. 
\begin{theorem}\label{thm:quantum-functions-NP} 
For exact function calculation with the use of an oracle in  $\C{NP}$ 
it holds that
  \begin{enumerate} 
  \item $\C{FP}_{||}^{\C{NP}[2^{k+1}-2]} 
    \subseteq  
    \C{FEQP}^{\C{NP}[2k]}$ for any $k\geq 0$. \label{thm:functions-k} 
  \item $\C{FP}_{||}^{\C{NP}} \subseteq \C{FEQP}^{\C{NP}[O(\log n)]}$. 
  \end{enumerate} 
\end{theorem} 
\begin{proof} 
Fix $k\geq 0$, the input $z$ of length $m$ and let $g$ be the function 
in $\C{FP}_{||}^{\sat [2^{k+1}-2]}$. Suppose that 
$g(z) = (a_1\cdots a_n) = a$ with $n = m^c$ for some $c$ depending on 
$g$. The goal is to obtain the following state: 
\begin{eqnarray} \label{eq:goal}
\ket{\textrm{Output}} & = & 
\frac{1}{\sqrt{2^n}}\sum_{x\in\{0,1\}^n}{{(-1)}^{\inner{x}{a}}\ket{x}}.
\end{eqnarray} 
Since with this state one application of $H^{\ox n}$ will give us 
$a = g(z)$ (see Equation~\ref{eq:hadamard-on-superposition}). 
Similar to 
Equation~\ref{eq:application-function} we can obtain this state if we had 
access to a function $f$ with the same property as the one in 
Equation~\ref{eq:special-f}.  
 
The goal thus is to transform the function we \emph{have} access to---$\sat$ 
in our case---into one that resembles the one in Equation~\ref{eq:special-f}. 
The way to do this is to make use of a quantum subroutine. Observe the 
following: the binary function $f_z(x) = \inner{x}{a}$ is  in 
$\C{P}_{||}^{\sat [2^{k+1}-2]}$ because we can first compute $g(z)= a$ 
with $2^{k+1}-2$ queries to $\sat$ and then determine $\inner{x}{a}$. 
By Theorem~\ref{thm:quantum-bin-search} this function is computable in 
$\C{EQP}^{\sat [k]}$. Hence, when we use this adaptive $\C{EQP}$ algorithm 
in superposition we have the desired function $f$. There is however one 
problem with this approach. The algorithm that comes out of  
Theorem~\ref{thm:quantum-bin-search} leaves several of the registers 
in states depending on the input $x$ and $\sat$. For example the algorithm 
that computes the parity of two function calls in one generates a phase  
of $(-1)$ depending on the value of the first function call (see 
Equation~\ref{eq:phase}). These changes in 
registers and phase shifts obstruct our base quantum machine and as a 
consequence the sum computed in 
Equation~\ref{eq:hadamard-on-superposition} does not work out the way 
we want (i.e., the interference pattern is different and terms do not
cancel out as nice as before.) 
 
The solution to this kind of `garbage' problem is as follows: 
\begin{enumerate} 
\item{Compute $f_z(x)$ with $k$ queries to $\sat$.}
\item {Copy the outcome onto an extra auxiliary qubit (by setting the 
  auxiliary bit $b$ to the \textsc{exclusive or} of $b$ and the outcome).} 
\item {Reverse the computation of $f_z(x)$ making another $k$ queries 
  to $\sat$.} 
\end{enumerate} 
Observe that when we compute $f_z(x)$ in this way, all the phase 
changes and registers are reset and are in the same state as before 
computing $f$, except for the auxiliary qubit that contains the 
answer. Since the subroutine was exact (i.e., in $\C{EQP}$) the answer 
bit is a classical bit and will not interfere with the rest of the 
computation. Note (see Section~\ref{sec:prel}) that this corresponds 
exactly to one oracle call to $f$. Thus we simulated $1$ call to $f$ 
with $2k$ queries to $\sat$ and hence have established a way
of producing the desired state of Equation~\ref{eq:goal}.

The second part of the theorem is proved in a similar way now using 
part 2 of Theorem~\ref{thm:classical-relation-sets}.  
\end{proof} 
 
\subsection{Terseness, and other complexity classes} 
The quantum techniques described above are quite general and can be 
applied to sets outside of $\C{NP}$. Classically the following 
question has been studied (see~\cite{Beigel88} for 
more information). For any set $A$ define the function 
$F_n^A(x_1,\ldots,x_n) = 
(A(x_1)\cdots A(x_n))$ which is an $n$ bit vector 
telling which of the 
$x_i$'s is in $A$ and which ones are not. A basic question now is: 
how many queries to $A$ do we need to compute $F_n^A$?  Sets for 
which $F_n^A$ can not be computed with less than $n$ queries to $A$ 
(i.e., $F_n^A \not \subseteq \C{FP}^{A[n-1]}$) are called \emph{$P$-terse.}  
We call the decision problem $A$ \emph{$P$-superterse} 
if $F_n^A \not \subseteq \C{FP}^{X[n-1]}$ for any set $X$. 
The next theorem shows that 
the notion $P$-superterse is not useful in the quantum setting. 
 
\begin{theorem}\label{thm:quantum-superterse} 
For any set $A$ there exists a set $X$ such that for all $n$ we have $F_n^A 
\subseteq \C{FEQP}^{X[1]}$.
\end{theorem} 
\begin{proof} 
Let $X$ be the following set: $X = \{\pair{z_1\cdots z_n,x}|
\inner{F_n^A(z_1,\ldots,z_n)}{x} \equiv 1 \bmod 2\}$. 
Using the the same approach as the proof of 
Theorem~\ref{thm:quantum-functions-NP} it is not hard to see that 
$F_n^A$ can be computed relative $X$ with only a single query. 
\end{proof} 
 
Using the same idea we can prove the following general theorem about
oracles for complexity classes other than $\C{NP}$. 
\begin{theorem} \label{thm:functionclasses}
Let $\mathcal{C}$ be a complexity class and the set $A$ 
  $\many$-complete for $\mathcal{C}$. 
\begin{enumerate} 
\item If $\mathcal{C}$ is closed under $\Tu$-reductions then 
$\C{FP}^{\mathcal{C}} = \C{FP}^{A} \subseteq 
\C{FEQP}^{A[1]} = \C{FEQP}^{\mathcal{C}[1]}$.
\item If $\mathcal{C}$ is closed under $\Tt$-reductions 
then 
$\C{FP}_{||}^{\mathcal{C}} = \C{FP}_{||}^{A} 
\subseteq 
\C{FEQP}^{A[1]} = \C{FEQP}^{\mathcal{C}[1]}$.
\end{enumerate} 
\end{theorem} 
\begin{proof} 
Let $f$ be the function we want to compute relative to $A$. Without 
loss of generality we assume that $\lng{f(z)} = \lng{z}^c$ for some 
$c$ depending on $f$. 
As before we construct the following set: 
\begin{eqnarray*} 
 X & = & \{\pair{z,y} \mathrel{|} \inner{f(z)}{y} \equiv 1 \bmod 2 
\mbox{, and $\lng{y} = \lng{z}^c$}  \}.
\end{eqnarray*} 
As in Theorem~\ref{thm:quantum-superterse} it follows that $f(z)$ is 
computable with one quantum query to $X$. Since ${\cal C}$ is closed 
under $\Tu$-reductions and $X \Tu A$, it follows that 
$X \in {\cal C}$. Furthermore, since $A$ is $\many$-complete for 
$\mathcal{C}$ it also follows that $X \many A$. Thus the quantum query 
can be made to $A$ itself instead of $X$. 
 The proof of the second part of the theorem is analogous to the first. 
\end{proof} 

This last theorem gives us immediately the following two corollaries 
about quantum computation with oracles for some known complexity classes.
\begin{corollary} 
\begin{eqnarray*}
\C{FP}^{\C{PSPACE}} & \subseteq & \C{FEQP}^{\C{PSPACE}[1]}\\
\C{FP}^{\C{EXP}} & \subseteq & \C{FEQP}^{\C{EXP}[1]}\\
\C{FP}^{\Delta_i^p} & \subseteq & \C{FEQP}^{\Delta_i^p[1]}
\end{eqnarray*} 
for the Delta levels $\Delta_i^p$ in the polynomial time hierarchy. 
\end{corollary} 
\begin{corollary} 
\begin{eqnarray*}
\C{FP}_{||}^{\C{PP}} & \subseteq &  \C{FEQP}^{\C{PP}[1]} \\ 
\C{FP}_{||}^{\Theta_i^p} & \subseteq & \C{FEQP}^{\Theta_i^p[1]}
\end{eqnarray*}
with $\Theta^p_{i+1}=\C{P}_{||}^{\Sigma_i^p}$. 
\end{corollary} 
The first corollary holds in particular for $A = \qbf$ 
(the set of true quantified Boolean formulae) which is $\C{PSPACE}$-complete.
 Observe also that the situation is quite 
different in the classical setting, since for $\C{EXP}$-complete sets 
the above is simply not true. 
 
\section{Conclusions and Open Problems} 
We have combined techniques from complexity theory with some of the 
known quantum algorithms. In doing so we showed that a quantum 
computer can compute certain functions with fewer queries than classical 
deterministic computers.  Many question however remain. Is it possible 
to get trade-off results between the adaptive class 
$\C{EQP}^{\C{NP}[k]}$ and the non-adaptive 
$\C{EQP}^{\C{NP}[2^k-1]}_{||}$ for quantum 
machines? Are the results we present here optimal? 
(Especially the recent results on exact searching in an ordered 
list\cite{FarhiGGS99} deserve further analysis as they seem to 
suggest a reduction of the quantum query complexity of 
Theorems~\ref{thm:quantum-bin-search} and \ref{thm:quantum-functions-NP} 
by a factor of two.)

What can one deduce 
from the assumption that $\C{P}^{\C{NP}} \subseteq \C{EQP}^{\C{NP}[1]}$?  
Is it true that for any set $A$ we have $\C{P}^A \subseteq \C{EQP}^{A[1]}$ 
or are there sets where this is not true? A random set would be a good 
candidate where more than $1$ quantum query is necessary. 

\section*{Acknowledgements} 
We thank Hein R\"ohrig and Leen Torenvliet for helpful proof reading
and Sophie Laplante for technical consultation. 
W.v.D. was supported by the European TMR Research Network ERP-4061PL95-1412,
Hewlett Packard, and the Institute for Logic, Language, and Computation
in Amsterdam.

\bibliographystyle{latex8} 

\end{document}